# Graphene-TMD van der Waals Heterostucture Plasmonics

*Partha Goswami and U.P.Tyagi*

*D.B.College, University of Delhi, Kalkaji, New Delhi-110019, India*
*physicsgoswami@gmail.com*; uptyagi@yahoo.co.in

**Abstract** The collective excitations of electrons in the bulk or at the surface, viz. plasmons, play an important role in the properties of materials, and have generated the field of "plasmonics." We report the observation of a highly unusual plasmon mode on the surface of Van der Waals heterostructures (vdWHs) of graphene monolayer on 2D transition metal dichalcogenide (Gr-TMD) substrate. Since the exponentially decaying fields of surface plasmon wave propagating along interface is highly sensitive to the ambient refractive index variations, such heterostructures are useful for ultra-sensitive bio-sensing.

**Main Text**

The local variation in charge density in materials gives rise to an electric field. The quanta of in-phase longitudinal density oscillation of charge carriers at the surface of materials driven by this electric field are defined as the surface plasmons (SPs). Inside the bulk, SPs evanesce severely owing to the heavy energy loss. These collective density oscillations can be excited in the conventional metal surfaces. They, however, suffer large energy losses, such as radiative loss and Ohmic loss and have bad tunability in a device[1]. The efficient wave localization up to mid-infrared and THz frequencies, the enhanced carriers density control by electrical gating and doping[2], the low losses, and the two dimensional (2D) nature of the collective excitations leading to stronger SP incarceration compared to those in metals make graphene to be a promising plasmonic material. Since the exploration of new materials for enhanced control and manipulation is an unceasing quest, in this report we focus on highly unusual plasmon mode, offering even more promising prospects for plasmonic applications, on the surface of Van der Waals heterostructures (vdWHs) of graphene monolayer on 2D transition metal dichalcogenide (Gr-TMD) substrate. Owing to its remarkable physical properties, such as high carrier mobility and electrostatically tunable optical properties, the hybrid structure is useful for manipulating electromagnetic signals at the deep-sub-wavelength scale. The surface plasmons can propagate sufficiently long distance along the surface in such structures under suitable condition, such as the increase of doping concentration in the Gr-TMD system.

The existence of quantum spin Hall (QSH) effect for pure graphene when moderate to large spin-orbit coupling (SOC) is taken into account was first suggested by Kane and Mele **[3,4]**. Due to the hybridization of the carbon orbitals with the d-orbitals of the transition metal, in Gr-TMD system sub-lattice-resolved, giant intrinsic SOCs ($V_{soc}^A, V_{soc}^B$) are possible. In fact, the strong SOCs originate from the d-orbitals of the transition metal atom. The other substrate-driven interactions**[5,6,7]** (SDIs) correspond to the same sub-lattice transfer of the electronic charge from Gr to TMD leading to a staggered potential gap($G$) without spin-flip, and the extrinsic Rashba spin-orbit coupling (RSOC) ($\zeta_{Rashba}$) that allows for the external tuning of the band gap in Gr-TMD and connects the nearest neighbors with spin-flip. Since the Rashba coupling in graphene is constant for mass-less Dirac electrons and does not depend on the momentum components ($k_x$, $k_y$), the intrinsic RSOC, which is modeled as $\alpha (k_y\sigma_x - k_x\sigma_y)$ (where $\sigma$'s are the Pauli matrices) in conventional semiconducting 2D electron gases, have not been considered here.

In order to set the stage, we choose a representation involving the states ($|c_{k,\tau,\uparrow}\rangle$, $|c_{k,\tau,\downarrow}\rangle$), where where $c_{k,\tau,s} = a_{k,\tau,s}$ ($b_{k,\tau,s}$) is the fermion annihilation operator for the momentum($k$)-valley($\tau$) - spin ($s$) state corresponding to the sub-lattice A(B), as our basis. In the basis chosen, including the interactions mentioned above, the effective, low-energy, dimensionless continuous model Hamiltonian [5,6,7,8] for Gr-TMD system (the wavevector components $ak_x$ and $ak_y$ are measured from the Dirac points $K$ and $K'$ and $a = 2.5$ A° is the lattice constant of graphene) may be written as $H_0 = [s_0 \otimes (\tau ak_x\sigma_x + ak_y\sigma_y) + G\ s_0 \otimes \sigma_z + \zeta_{Rashba}\ (\tau\sigma_x \otimes s_y - \sigma_y \otimes s_x) + \tau(V^A_{soc}\sigma_+ + V^B_{soc}\sigma_-) \otimes s_z]$ where the valley index $\tau$ is +1(−1) for the point $K$ ($K'$) and $\sigma_\pm = \frac{1}{2}(\sigma_z \pm \sigma_0)$. The Pauli matrices $\sigma_i$ and $s_i$ respectively, are affiliated with the pseudo-spin, and the real spin of the Dirac electronic states. The terms present in the Hamiltonian are made dimensionless dividing by the energy term ($\hbar v_F/a$). Here the nearest neighbor hopping is parameterized by a hybridization $t$ ( $\hbar v_F/a = (\sqrt{3}/2)t$). On a quick side note, we add that the vdW heterostructure under consideration, if modified slightly, could be used as devices other than a sensor. For example, if the graphene layer in the vdWH is further proximitized from the top with a magnetic material, such as monolayer ferromagnetic insulator $Cr_2Ge_2Te_6$, one needs to add $(M^A_{soc}\sigma_+ - M^B_{soc}\sigma_-) \otimes s_z$ to the Hamiltonian $H_0$, where $(M^A_{ex}, M^B_{ex})$ are the sub-lattice-resolved proximity exchange interaction parameters. Furthermore, if tunability is imparted via combined top and back electrostatic gates, one obtains an efficient spin-orbit torque operated field-effect transistor. The function of the gate voltage is to bring about changes in the chemical potential $\mu$ of the fermion number. In fact, the relation we have used here to tune the Plasmon frequency with gate voltage ( see Figure 1) is

$\mu \approx \varepsilon_a [(m^2 + \frac{2eV_g}{\varepsilon_a})^{\frac{1}{2}} - m]$, where $m$ is the dimension-less ideality factor, $V_g$ is the gate voltage and $\varepsilon_a$ is the characteristic energy scale.

The SDIs included in the Hamiltonian matrix $H_0$ lead to spin-valley resolved bands of the form $\varepsilon_{\tau,s,\lambda}$ ($k$,$M$)=[$sP(M, \tau) \zeta_{Rashba} + \lambda\sqrt{\{\varepsilon^2_k + Q^2(M,s,\tau)\}}$] for our massive Dirac system, where we have assumed $M = (M^A_{ex}, M^B_{ex})$ for simplicity and $Q(M,s,\tau)$ is the staggered mass-gap. The band structure consists of two spin-chiral conduction bands and two spin-chiral valence bands. Because of the spin-mixing driven by the Rashba coupling, the spin is no longer a good quantum number. Therefore, the resulting eigenstates may be denoted by the spin- chirality index $s = \pm 1$. Here $\lambda = +(-)$ indicates the conduction (valence) band; $\varepsilon_k$ is the spin-valley degenerate energy dispersion of the pristine, pure graphene. The band structure (i) preserves the inversion symmetry($\varepsilon_{\tau,s,\lambda}$ ($k$,$M$) = $\varepsilon_{\tau,s,\lambda}$(−$k$,$M$) ), (ii) breaks the TRS ($\varepsilon_{\tau,s,\lambda}$ ($k$,$M$) $\neq \varepsilon_{\tau,-s,\lambda}$ (−$k$,$M$) ), (iii) involves an effective Zeeman field $P(M, \tau) \zeta_{Rashba}$ due to the interplay of SDIs, and (iv) involves the spin-orbit interaction led avoided crossing of the bands with the same spins around Dirac points. The anti-crossing of the non-parabolic bands have been shown by MacDonald et al.[9] several years ago. Since with hole doping the Fermi surface of Gr-(TMD) shifts to a lower energy, as a consequence the inter-band transitions with transition energy below twice the Fermi energy become forbidden. It leads to a decrease in higher frequency inter-band (optical) absorption. At the same time, the lower frequency (far infrared and terahertz (THz))free carrier absorption (i.e. intra-band transition) increases dramatically. Therefore, the intra-band transitions are needed to be considered only for Gr-TMD. For these transitions, since the focus is on the long wavelength regime, the transitions between two Dirac nodes located at different momentum could be neglected. The spin-valley resolved bands $\varepsilon_{\tau,s,\lambda}$ together with these assumptions are the starting points to calculate Plasmon mode for our Gr-TMD system. Within the random phase approximation(RPA), by finding the zeros of the frequency($\omega$)-dependent dielectric function involving only the intra-band transition, we obtain a rather unusual Plasmon mode for a finite chemical potential. Unlike the pure graphene case, the corresponding dispersion relation is $\hbar\omega'_{pl} \sim A^{\frac{1}{3}}(aq)^{\frac{2}{3}}\left(ak_F(\mu',M)\right)$ where $A = \pi\left(\frac{\frac{e^2}{2a\varepsilon_0\varepsilon_r}}{\left(\frac{\hbar v_F}{a}\right)}\right)$, $\hbar\omega'_{pl} = \frac{\hbar\omega_{pl}}{\left(\frac{\hbar v_F}{a}\right)}$, $\varepsilon_0\varepsilon_r$ is the permittivity of the ambient medium, $q$ is wave vector, and $ak_F(\mu',M)$ is the dimensionless Fermi momentum. It is given by $ak_F(\mu',M) = \left(\frac{1}{4}\sum_{\tau,s}\sqrt{\{(\mu' - sP(M,\tau)\zeta_{Rashba})^2 - Q^2(M,s,\tau)\}}\right)$. The term $\mu' = \mu/\left(\frac{\hbar v_F}{a}\right)$. It is evident that the power law dependence of the plasmon frequency on wave vector is of the type $q^{2/3}$ for the gapped(massive) Dirac system, whereas for the ungapped graphene case it is $q^{1/2}$.

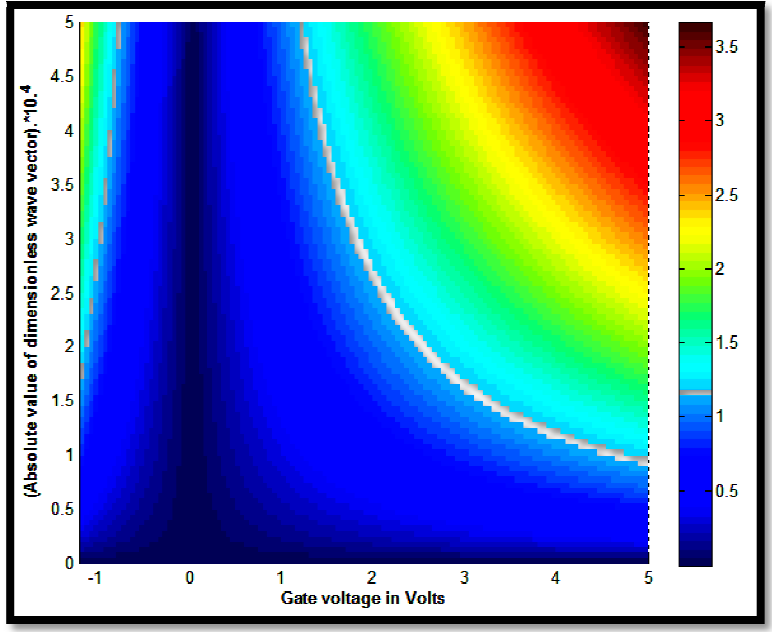

**Figure 1** A contour plot showing the Plasmon frequency as a function of the gate voltage ($V_g$) and the absolute value of dimensionless wave vector ($a|\boldsymbol{k}|$)in the case of graphene on WSe$_2$ at T ≈ 0 K. The plot and the colour-bar indicate the increase in the Plasmon frequency with the increase in the absolute value of the gate votage at a given wave vector. The enclosed region of ($a|\boldsymbol{k}|$) - $V_g$ plane does not correspond to much of an increment.

Furthermore, since $\omega'_{pl}$ depends on the Fermi momentum, an approximate dependence $(\hbar\omega_{pl}) \propto n^{1/2}$ is expected for Gr-TMD system ( and not $n^{1/4}$ as in the case of graphene), where $n$ is the carrier concentration. Regarding the response modulation, we find that whereas the tunability of graphene plasmons stems from the $\omega_{pl}^{Gr} \propto n^{1/4}$ ( or, $\omega_{pl}^{Gr} \propto \mu^{1/2}$) dependence albeit the electrostatic/ chemical doping, in the case of the Gr-TMD also it is due to the $\mu$ (or the gate voltage ) dependence. The presence of $(ak_F(\mu', M))$ term only alters the nature of the dependence. The graphical representation is shown in Figure 1. There is an increase in the Plasmon frequency with the increase in the absolute value of the gate votage at a given moderate to high value of the wave vector. In sharp contrast to ordinary plasmon modes, this mode exhibits quasi-linear dispersion into the second Brillouin zone and remains prominent with remarkably weak damping not seen in any other systems.

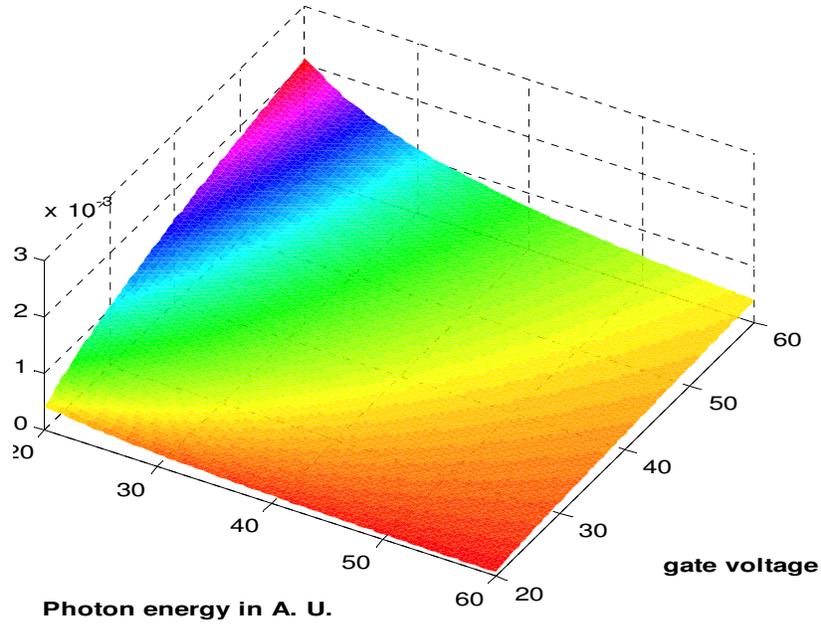

**Figure 2**. A 3 D plot of the FWHM as a function of the photon energy and the gate voltage. The full width at half maximum is narrow at high photon energy and low gate voltage.

A potential application of the graphene-transition metal dichalcogenide (TMDC) hybrid structures is the plasmonic meta-surface for ultra-sensitive bio-sensing. The sensing principle is the utilization of the exponentially decaying fields of a surface plasmon wave (SPW) propagating along interface, which is highly sensitive to the ambient refractive index variations, such as induced by bio affinity interactions at the sensor surface. An important sensor performance parameter is the reciprocal of full width at half maximum (FWHM). Therefore, for a SPW sensor with excellent performance the FWHM should be as small as possible. Close to the Plasmon frequency $\omega_{pl}$, the spectral function may be represented as a Lorentzian in $(\omega - \omega_{pl})$ with the full width at half maximum and height, respectively, given by two different functions of $(\omega, \mu, q)$. We note that FWHM could be controlled by changing the chemical potential through the electrostatic / chemical doping and could be made as narrow as is required. Furthermore, we find that the full width at half maximum is narrow at high photon energy and low gate voltage (see Figure 2). It may now be noted that though metal-dielectric based surface Plasmon (SP) resonance has been widely employed for sensing applications **[10,11]**, such as gas sensing, temperature sensing, and bio-sensing, during the last two decades due to its high sensitivity and reliability, a SP sensor based on the Gr-TMD heterostructure, however, has greater advantage as it is likely to have good (tunable) performance. In conclusion, the Gr-TMD plasmons (THz varieties) have unusual properties and offer promising prospects for plasmonic applications. It is useful for manipulating electromagnetic signals at the deep-sub-wavelength scale owing to its remarkable physical properties, e.g., high carrier mobility and electrostatically tunable optical properties. Thus, the extraordinary properties of SPs in Gr-TMD, plus its good flexibility, and stability make it a good candidate for varieties of applications, including THz technology, energy storage, biotechnology, medical sciences, electronics, optics, and so on.